\input psfig
\documentstyle[twoside,fleqn,espcrc2]{article}

\newcommand{\AmS}{{\protect\the\textfont2
  A\kern-.1667em\lower.5ex\hbox{M}\kern-.125emS}}

\title{The $\eta'$ propagator in quenched QCD\thanks{Talk presented by H. Thacker}}

\author{W. Bardeen,\address{Fermilab, P.O. Box 500, Batavia, IL 60510}%
        A. Duncan,\address{Dept. of Physics and Astronomy,University of Pittsburgh, Pittsburgh, PA 15260}%
        E.~Eichten,$^{\rm a}$
        S.~Perrucci,$^{\rm c}$
        and 
        H.~Thacker\address{Dept. of Physics, University of Virginia, Charlottesville, VA 22901}}

\begin{document}

\begin{abstract}
 
The calculation of the $\eta'$ hairpin diagram is carried out in
the modified quenched approximation (MQA) in which 
the lattice artifact which causes exceptional configurations is
removed by shifting observed
poles at $\kappa<\kappa_c$ in the quark propagators to the critical value of hopping
parameter. By this method, the $\eta'$ propagator 
can be accurately calculated even for very light quark mass.
A determination of the topological susceptibility for quenched
QCD is also obtained, using the fermionic method of Smit and Vink 
to calculate winding numbers. 

\end{abstract}

\maketitle

\section{INTRODUCTION}

The properties of
the flavor singlet pseudoscalar $\eta'$ meson provide a unique phenomenological
window on the topological structure of QCD. The mass of the $\eta'$
is believed to arise from topologically nontrivial gauge configurations
via the axial $U(1)_A$ anomaly. In a semiclassical treatment, the
effect of instanton contributions to the $q\bar{q}$ annihilation 
(``hairpin'') diagram breaks the degeneracy between the $\eta'$
and the flavor octet Goldstone bosons. In an effective chiral
lagrangian description of QCD, the $q\bar{q}$ hairpin can be 
interpreted as an $\eta'$ mass insertion. Within the large $N_c$
approximation, Witten and Veneziano \cite{WV} showed that the 
topological theory of $U(1)_A$ breaking is consistent with 
the mass-insertion view, and that in this approximation, the
$\eta'$ mass in the chiral limit 
is proportional to the topological susceptibility
$\chi_t$ of {\it pure gauge} (quenched) QCD. 
\begin{equation}
m_0^2 = \frac{2N_f}{f_{\pi}^2}\chi_t
\end{equation}
(Note: Here we take $f_{\pi}$ normalized to have the experimental value
of $\approx 96$ MeV. This differs from that used in Ref. \cite{Lat96}
by a factor $\sqrt{2}$.)

The study of the $\eta'$ propagator in lattice
QCD is of great interest, not only as a quantitative check of
the theory of the $U(1)$ problem, but also as a particularly
sensitive probe of the differences between quenched and full
QCD. For example, if the hairpin diagram behaves like a mass
insertion, the quenched $\eta'$ momentum-space propagator is expected
to include a double Goldstone pole contribution. 
\begin{equation}
\label{eq:doublepole}
\Delta_h(p^2) \propto \frac{1}{(p^2+m_{\pi}^2)}m_0^2
\frac{1}{(p^2+m_{\pi}^2)}
\end{equation}
As a result,
quenched chiral logs arising from virtual $\eta'$ loops 
will complicate the chiral behavior of quenched QCD compared with
that of the full theory\cite{Sharpe}. One of the purposes of the study 
reported here is to investigate in detail the time-dependence of
the hairpin contribution to the $\eta'$ propagator and compare
it with that expected from the double pole structure (\ref{eq:doublepole}).

The method we use to calculate closed loops which originate
at a given site of the lattice was introduced by 
Kuramashi, et al \cite{Kuramashi} in their original study of
the $\eta'$ mass. In this ``allsource'' technique, the 
quark propagator is calculated using a source given by a unit
color-spin vector on {\it every} space-time point of the lattice.
The closed quark loop from a given point is then calculated
by assuming random-phase cancellation of other non-gauge-invariant
terms. 

\begin{figure}
\vspace*{4.6cm}
\includegraphics{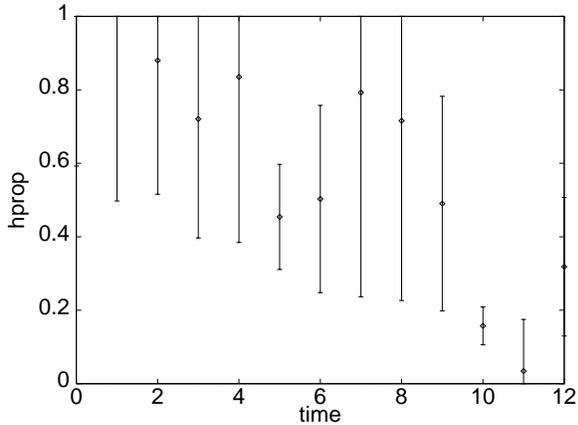}
\vspace{0.5cm}
\caption[]{The unimproved hairpin propagator on a $12^3\times 24$
lattice at $\beta=5.7$ and $\kappa=.1675$ }
\label{fig:hp_raw}
\end{figure}
A particular difficulty encountered in the $\eta'$ hairpin
calculation is the presence of rapidly increasing non-gaussian
errors in the limit of small quark mass due to exceptional
configurations. This problem is more severe in the hairpin
calculation than it is in ordinary hadron spectrum calculations.
Recently, the origin of the exceptional configuration problem
has been traced to the presence of topological zero mode
poles in the quark propagator which have been displaced to
values of the hopping parameter below $\kappa_c$ by lattice effects.
\cite{Eichten} A practical method for removing this
lattice artifact, the modified quenched approximation (MQA),
has been proposed in Ref. \cite{Eichten}. 
An example of the MQA improvement of the hairpin
propagator is shown in Figs. 1 and 2. The propagators
shown in Figs. 1 and 2 were obtained on a $12^3\times 24$ lattice
at $\beta=5.7$ with a hopping
parameter of $\kappa=.1675$ ($m_q\approx 38$ MeV). The improvement
of errors due to the MQA pole-shifting is impressive. For lighter
quark masses, the data for the $12^3\times 24$ lattice
without the MQA improvement is unusable
due to extremely large errors. 
After MQA improvement it is possible to accurately calculate the hairpin
diagram down to very light quark masses. 
In our calculations, we have shifted all poles 
which were found
below $\kappa =.1690$ and have calculated the hairpin for
$\kappa$ values up to .1685. It may be feasible to go to even
lighter quark mass values, but the calculation of pole residues
above $\kappa=.1690$ becomes rapidly more expensive in computer
time.

\begin{figure}
\vspace*{4.6cm}
\includegraphics{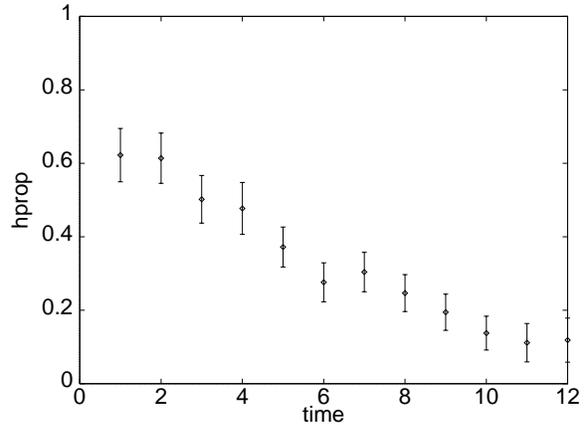}
\vspace{0.5cm}
\caption[]{The MQA improved hairpin propagator on a $12^3\times 24$
lattice at $\beta=5.7$ and $\kappa=.1675$ }
\label{fig:hp_mqa}
\end{figure}

\section{THE $\eta'$ PROPAGATOR}

The hairpin diagram has been calculated on two ensembles of
gauge configurations available in the ACPMAPS library. One
ensemble included 200 configurations on a $12^3\times 24$ lattice
at $\beta=5.7$. The other ensemble 
consists of 200 configurations
at $\beta=5.7$ on a $16^3\times 32$ lattice. All together we
have calculated the results for seven different values of
hopping parameter ranging from .161 to .1685. 

\begin{figure}
\vspace*{4.6cm}
\includegraphics{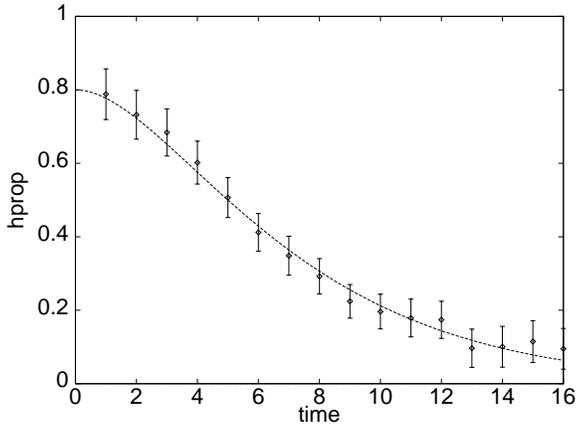}
\vspace{0.5cm}
\caption[]{The MQA improved hairpin propagator on a $16^3\times 32$
lattice at $\beta=5.7$ and $\kappa=.1680$. The solid line is the
time dependence of a pure double Goldstone pole, Eq.~(4) expected if the
$\eta'$ hairpin vertex is a simple mass insertion.}
\label{fig:effmass}
\end{figure}
In general, the hairpin insertion may not be a simple $p^2$-independent
mass insertion, but instead may exhibit some $p^2$-dependence.
Expanding around the pion mass shell, we may write
\begin{equation}
\label{eq:fullfit}
m_0^2 \rightarrow h(p^2) = m_0^2+\alpha(p^2-m_{\pi^2}) +\ldots
\end{equation}
The second term corresponds to an additon to the $\eta'$ kinetic
term in the chiral lagrangian.
The form of the hairpin insertion may be determined by studying
the time-dependence of the hairpin propagator. If the hairpin
vertex is a simple $p^2$-independent mass insertion, the measured
propagator at zero 3-momentum should behave according to a pure
double-pole formula,
\begin{eqnarray}
\label{eq:dptime}
\tilde{\Delta}_h(\vec{p}=0,t)= \frac{C}{4m_{\pi}^3}(1+m_{\pi}t)
\exp(-m_{\pi}t)\\ \nonumber
+(t\leftrightarrow (T-t))
\end{eqnarray}
The value of $m_{\pi}$ is determined quite accurately from the
valence pion propagator, so the only adjustable parameter in 
this fit is the overall normalization $C$. The second term in (3) would
contribute a single-pole term to the propagator (2), which gives
a term in (4) with pure exponential time dependence. Thus, both single and 
double pole terms are included by replacing the factor $C(1+m_{\pi}T)$
in (4) by $(C_1+C_2m_{\pi}t)$.
The prelimnary results of this analysis
indicate that there is very little $p^2$ dependence of the
hairpin insertion. Indeed, the time dependence of the
propagators is remarkably well described by the pure
double-pole formula (4) over the entire
observable range of time separations. An example for
the $16^3\times 32$ lattice at $\kappa=.1680$ is shown
in Fig. 2. 

By fitting the propagators to (\ref{eq:dptime}),
and dividing out apppropriate factors obtained from the valence
pion propagator, we obtain a value for $m_0$ at each $\kappa$.
The $16^3$ results are consistent
with those reported last year \cite{Lat96}. The finite volume increase 
observed in the $12^3$ data \cite{Lat96} is found to be largely an effect of having more
nearby visible poles on the smaller box. After the MQA shift, the results on the two
box sizes are within a standard deviation of each other. 
The MQA analysis of $f_{\pi}$ is not yet completed, but using
previous results for $f_{\pi}$, calculated on the $12^3\times 24$
configurations, 
we obtain an effective chiral log
parameter of $\delta\approx .04$ at $\kappa = .168$. 
The full results and comparison with other work will be presented
elsewhere.

As a byproduct of the hairpin propagator calculation, we may
calculate the integrated pseudoscalar charge density
$Q_5 = \int \bar{\psi}\gamma_5\psi\;d^4x$ on each lattice
in the ensemble. As first suggested by Smit and Vink \cite{Smit},
this provides a fermionic method for determining the winding
number $\nu$ of a gauge configuration, using the integrated
anomaly equation, which gives $\nu =\lim_{m\rightarrow 0} mQ_5$.
Using the allsource quark propagators, improved by the MQA
pole shifting procedure, we have employed the Smit-Vink method
to calculate the topological susceptibility
$\chi_t=\langle\nu^2\rangle/V$
(where $V$ is the lattice 4-volume) for both the $12^3\times 24$
and $16^3\times 32$ ensembles. The results show only a mild
$\kappa$ dependence and can be sensibly extrapolated to
zero quark mass. Using the scale $a^{-1}=1.15$ GeV taken from
charmonium, we obtain $\chi_t = 12.8\pm 1.4\times 10^{-4}$ GeV$^4$
for the $12^3\times 24$ lattice, and $\chi_t = 10.8\pm 1.2\times 10^{-4}$
GeV$^4$ for the $16^3\times 32$ lattice. This can be compared with
$11.5\times 10^{-4}$ GeV$^4$ from the WV formula (1).

\end{document}